\documentclass[aps,preprintnumbers,floats, prd,nofootinbib]{revtex4-1}

\usepackage{epsfig}
\usepackage{amssymb}
\usepackage{bm}
\usepackage{hyperref}
\usepackage{color}
\usepackage{amsmath}

\newcommand{\be}{\begin{equation}}
\newcommand{\ee}{\end{equation}}
\newcommand{\bea}{\begin{eqnarray}}
\newcommand{\eea}{\end{eqnarray}}

\newcommand{\nn}{\nonumber}

\newcommand{\degree}{^\circ}

\newcommand{\sci}[2]{#1$\times$10$^{\text{#2}}$}
\newcommand{\ck}[1]{\textcolor{black}{#1}}

\begin{document}

\title{Enhanced Gamma Ray Signals in Cosmic Proton-Wimp Collisions Due to Hadronization}

\preprint{IPPP/12/72, DCPT/12/144}

\author{Spencer~Chang$^{1}$, Yu~Gao$^{1}$, Michael~Spannowsky$^{2}$\\}

\vspace{0.5cm}
\affiliation{%
\bigskip
$^1$Institute of Theoretical Science, University of Oregon, Eugene, OR  97403-5203, USA \\
$^2$Institute for Particle Physics Phenomenology, Department of Physics, Durham University, DH1 3LE, United Kingdom\\}

\begin{abstract}
In this paper, we investigate the gamma ray signal produced from dark matter 
collisions with high energy cosmic protons.  
Notably, we extend past results by including important hadronization effects.  
Showering and hadronization produces a high multiplicity of photons from the decays of hadrons, whose rate is not suppressed by
the fine structure constant.   
Notably, proton remnants that
do not participate in hard scattering, can produce a large rate of photons in the forward direction.  
These effects significantly enhance the photon rate and alter the energy and angular distributions compared to previous results
which used only parton level calculations.  
Due to this modification, the gamma ray signal from the nearby active galactic nuclei Centaurus A 
is potentially testable in future Fermi-LAT and HESS measurements, for a dark matter
mass and coupling consistent with current XENON100 bounds.  
\end{abstract}

\maketitle

\section{Introduction}

The fundamental nature of dark matter remains mysterious to this day.  The impressive consistency with the dark matter paradigm on a wide range of scales is tempered by the fact that only its universal gravitational interactions have been observed.  To go further in our understanding, obtaining solid evidence of dark matter interactions with the Standard Model would be of enormous value.  One promising approach to test such interactions is observing cosmic gamma rays. Dark matter can produce gamma rays through annihilation or decay and is currently being searched for by experiments such as Fermi-LAT~\cite{Falcone:2010fk} and HESS~\cite{Raue:2009vp}.  

Another dark matter gamma ray signal that is less well known is gamma rays produced in collisions of dark matter with high energy cosmic particles.  This indirect signal was proposed initially by Ellliot and Wells \cite{Bloom:1997vm} and has been reinvestigated more recently.    Specifically, gamma rays from collisions between cosmic ray (CR) and dark matter (DM) particles have been studied in  regions with concentrated dark matter and high energy cosmic ray 
flux, near the center of active galactic nuclei (AGN)~\cite{Gorchtein:2010xa, Huang:2011dg} and also the central region of the Milky Way~\cite{Profumo:2011jt}.
Refs.~\cite{Bloom:1997vm, Gorchtein:2010xa, Huang:2011dg} 
investigated the parton-level radiation and showed that the scattering between 
dark matter and the cosmic electrons in AGNs can be a promising gamma ray signal.  On the other hand, \cite{Gorchtein:2010xa} found that proton-wimp interactions lead to a less promising signal due to the quark's fractional charge and momentum distribution in the proton.  However, this analysis was a parton-only calculation and neglects  important effects, such as showering and hadronization.     
In this study, we demonstrate that the hadronic shower contributes a large number of photons from hadron decays.  Notably, this photon production is enhanced relative to the parton-only calculation due to large multiplicity and the lack of suppression by $\alpha_{QED}$.
Furthermore, these photons have a substantially altered energy spectra, with energies extending to higher values compared to those from the hard scattering process. 
The hard scattering occurs at an energy scale related to the dark matter mass, which can be much lower
than the total incoming proton energy. The showering from the rest of the proton, which does not participate
in the hard scattering and is likely to carry the majority of the  incoming energy, can emit very energetic photons
in the forward direction.  For AGNs, this enhancement allows the small fraction of protons directed towards the Earth to give 
a significant gamma ray contribution.  
To summarize, we find that the photons of the shower are an important modification to the gamma ray signal from proton-dark matter scattering which greatly enhances 
the rate and modifies the shape of the energy spectrum.

In our analysis we adopt a toy model of a Majorana fermion dark matter,
that couples to the right-handed up quark through a heavy charged scalar.
This simple model serves as a template for models where the dark matter - cosmic ray
collisions give an important photon signal, while other signals, like dark matter 
annihilation, become suppressed.  
With a heavy partner to a standard model quark, the collision
process can undergo an s-channel resonance, leading to a large scattering cross-section.  This can be consistent with bounds from direct detection experiments where the $\sim$keV recoil energy is much lower 
than the resonance energy. Interestingly, as most of the photon radiation is emitted on resonance,
the photon spectral structure is determined by the mass difference between the DM and
the heavy $p/e^-$ partner, instead of the mass of DM itself. 
This creates freedom in the signal's energy scale that differs from that from DM
annihilation and decay cases.

The rest of this paper is organized as follows.  In Section~\ref{sect:theory} we discuss the 
resonant scattering process in the toy model.
We take into account the allowed mass range from the latest XENON100 constraints~\cite{Aprile:2012nq}, to determine allowed signal benchmarks.   Section~\ref{sect:spectrum} outlines the calculation of the gamma ray spectrum.
In Section~\ref{sect:AGN} we study the signal from collisions off cosmic rays for the AGN Centaurus A. In Section~\ref{sect:diffuse} we comment on the enhancement on the gamma rays from dark matter collisions with diffuse cosmic rays.  A summary is presented in Section~\ref{sect:summary}.  Finally, in the Appendix~\ref{app:source_spec}, we list some important formulas for diffuse cosmic ray scattering.

\section{Resonant Cosmic Ray-Dark Matter scattering}
\label{sect:theory}

\begin{figure}[h]
\includegraphics[scale=0.6]{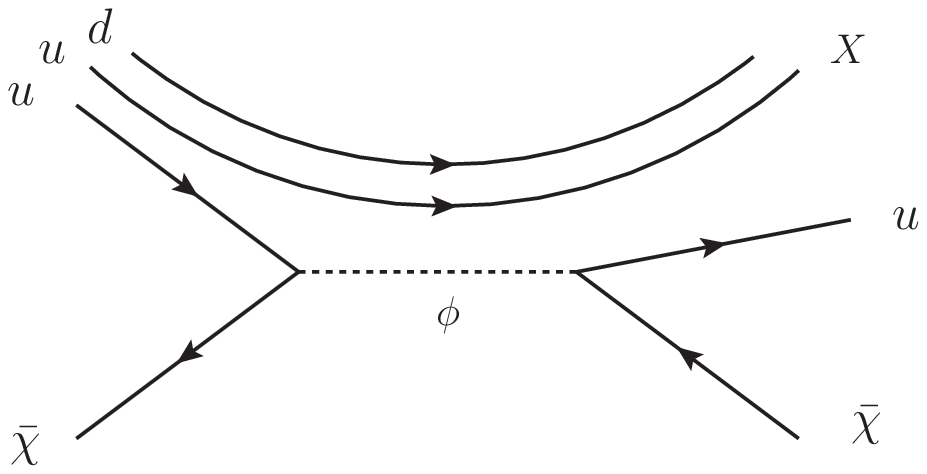}
\includegraphics[scale=0.6]{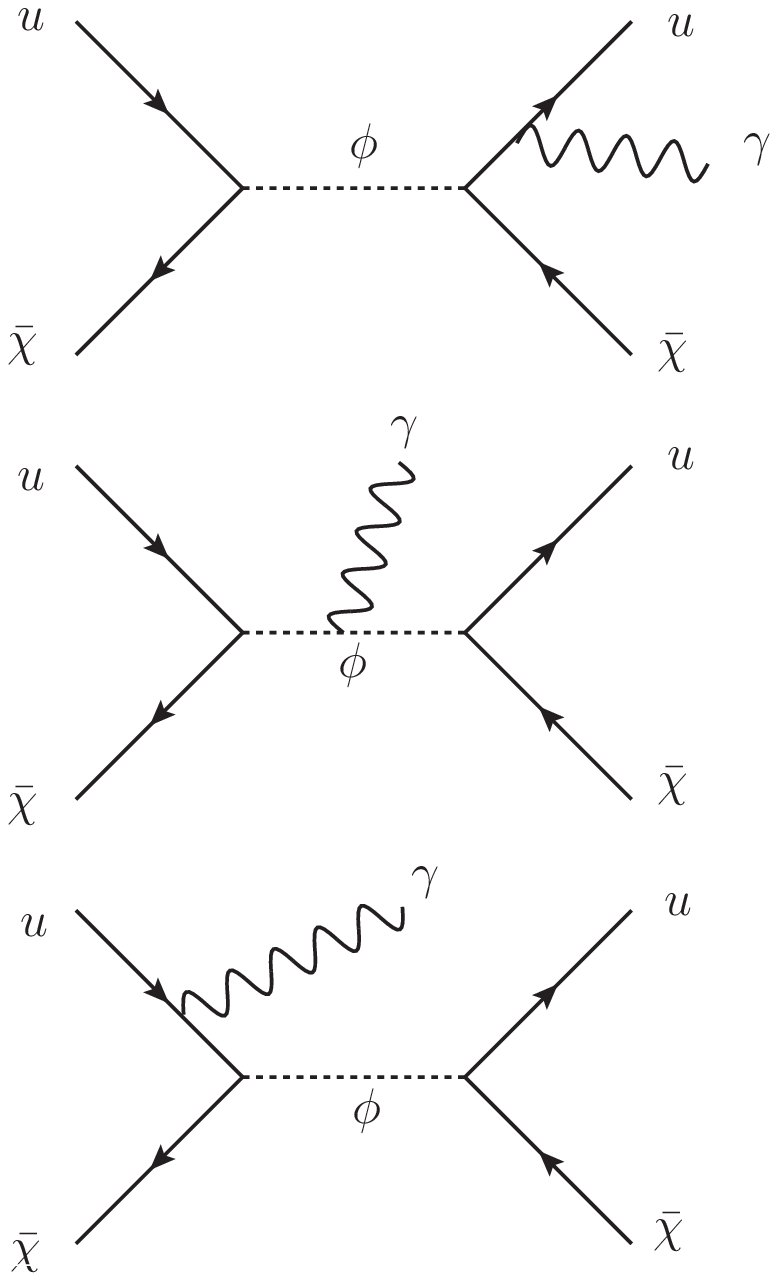}
\caption{Resonant s-channel diagrams. The charged $u$-partner, $\phi$, can also radiate photons. The left panel is the 
leading 2 to 2 collision and the right panel shows the leading $\chi u\rightarrow \chi u\gamma$ processes
that dominates the hard event's photon emissions.}
\label{fig:feynman_diagrams}
\end{figure}

The rates of cosmic ray-dark matter scattering production of gamma rays are particularly
interesting when there is an $s$-channel resonance enhancement \cite{Gorchtein:2010xa}.  
For example,  a heavy scalar partner to the up  quark, $\phi$, can mediate an s-channel 
resonance as shown in the left column of
Fig.\ref{fig:feynman_diagrams}.  This should be compared with 
the three processes in the right column which illustrate the leading 
parton-level photon emission.   

In the galaxy, the dark matter $\chi$ is non-relativistic.  Thus, the condition on the up-quark's energy to hit the resonance is (neglecting the up quark mass)
\bea
M_{\phi}^2 = s= (p_{\chi}+ p_{u})^2 = M_{\chi}^2 + 2 E_u M_\chi
\hspace {1cm} \text{or}\hspace{1cm} E_u=E_{res}&\equiv&\frac{M_{\phi}^2-M_{\chi}^2}{2M_{\chi}}
\label{eq:res}
\eea
where $s$ is the square of the center-of-mass (COM) frame energy. 
To realize this signal, we will consider a theory with a Majorana fermion dark matter with the interaction,
\be 
\mathcal{L}_{int}=y \, \bar{\chi}P_R u\, \phi^* + h.c.
\ee
where the dark matter $\chi$ only couples right-handedly to the up quark via the scalar $\phi$,
a colored partner to the up quark that carries the same electric charge.
We choose the dark matter to be Majorana to avoid
inducing large spin-independent scattering which would be strongly excluded by direct detection experiments.
Furthermore, the right-handed coupling ensures that  non-relativistic annihilation $\chi \chi \to u\bar{u}$ is
chirality suppressed. 
For proton energies above $E_{res}$, integrating out the parton distribution function (PDF) always ensures hitting the s-channel
resonance.  Due to the enhancement at the resonance,  the total high energy scattering cross-section increases as $\sim y^2$.  As we will show,  the direct detection bounds are more stringent at larger coupling, which can be avoided by taking a  larger mass gap between $\chi$ and $\phi$.  However, since this mass gap also determines the gamma ray spectra, there is a complicated interplay between satisfying direct detection limits and enhancing the gamma ray signal.    
Thus, to be concrete, we set $y=1$ throughout this paper and will choose the mass gap to be consistent with direct detection limits.

The $\phi$ decay width is
\be 
\Gamma_{\phi}=\frac{y^2}{16\pi}M_{\phi}\left(1-\frac{M_\chi^2}{M_\phi^2} \right)^2
\ee
which is less than ${\cal O}(10^{-2} M_{\phi})$ in our analysis. For such a narrow width, 
the s-channel resonance dominates when
kinematically allowed and
the total $\sigma_{\chi p}$ shoots up for $E_p>E_{res}$.
While $\sigma_{\chi p}$ continues to grow with proton
energy, the cosmic proton flux normally decreases as a power-law spectrum. Thus the total gamma ray signal 
depends on the energy where  $\sigma_{\chi p}$ turns up, 
which is determined by the mass gap between $\chi,\phi$.  

\begin{figure}[h]
\includegraphics[scale=0.6]{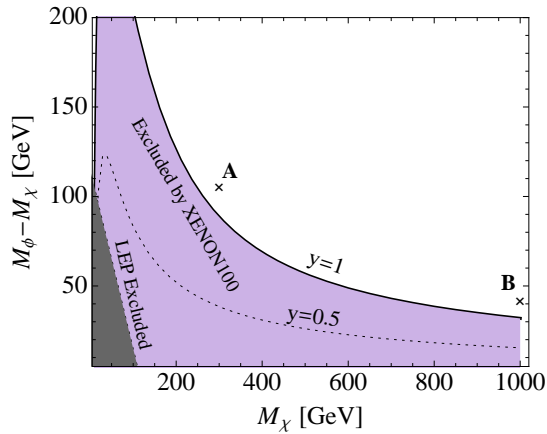}
\caption{Dark matter and u-partner masses allowed by 2012 XENON100 results at 90\% C.L.
\ck{Sample mass points are marked above the $y=1$ bound. The XENON100
constraint with $y=0.5$ (dotted) is also shown as the dotted curve for comparison.}}
\label{fig:masses}
\end{figure}

This mass gap has a lower bound from direct detection \ck{experiments} since the scattering 
rate is enhanced in the squeezed limit \ck{\cite{Hisano:2011um}}.  \ck{In Fig.~\ref{fig:masses}, we show the minimal mass differences allowed by recent XENON100
results~\cite{Aprile:2012nq} \ck{at 90\% confidence level}.}  
We have included the spin-dependent (SD) scattering and also the spin-independent (SI) scattering induced by the twist-2 operator \cite{Drees:1993bu}, which are comparable in importance near the bound.  Due to the resultant change in the energy spectrum, we cannot use XENON100's limit which is based on a profile likelihood.  To construct our own limit, we use XENON100's hard discrimination cut acceptance shown in their Fig.~1 \cite{Aprile:2012nq} and require less than 5.3 expected signal events, which is the 90\% CL limit given their two observed events.   As a cross check, our derived limit on the SI cross section $\sigma_{\chi N}$ is slightly weaker than their observed profile likelihood limit, but consistent within the 1$\sigma$ expected sensitivity band in their Fig.~3 \cite{Aprile:2012nq}. We refer to ~\cite{Hisano:2010ct, Hisano:2011um} for further details of the SD and SI cross section
calculations.

As two benchmarks, we use points  
A and B at $(m_\chi,m_\phi)=$(300, 405) GeV and (1, 1.04) TeV, 
While a narrower mass gap can be allowed for a lower $y$, the scattering cross-section
decreases faster than the gain from a lower resonance CR energy 
(for the assumed $E^{-2}$ spectrum in this paper).  
Thus the gamma ray signal  turns out to be less favorable with smaller couplings.
At the sample points, non-relativistic annihilation $\chi\chi\rightarrow u\bar{u}$ is 
chirality suppressed by the small u-quark mass.  
The leading annihilation process is 
$\chi\chi\rightarrow u\bar{u}g$, with a sub-picobarn $\left<v\sigma\right>$ 
that is allowed by PAMELA~\cite{Adriani:2010rc} results on the local $\bar{p}/p$ ratio. 

Although a light dark matter at multi-GeV mass may evade the direct detection bounds \cite{lightdm},
in our SUSY-inspired toy model the charged scalar partner $\phi$ is required to have a very small mass difference from
the dark matter candidate, in order avoid detection at LEP and LHC. While such a scenario
is possible, it is beyond the scope of this paper to fully study the collider bound at a narrow corner of our toy model.

\section{Prompt gamma spectrum}
\label{sect:spectrum}

When cosmic rays collide off dark matter, photons are emitted either 
directly from the hard scattering process or produced during the shower.
The parton level spectrum has been studied in detail in Ref.~\cite{Gorchtein:2010xa}.
The high energy photons are emitted in different ways during $\chi,e^-$
and $\chi,p$ collisions, as described below.

In the case of $\chi,e^-$ collision, the leading photon emission is 
through the $\chi e^-\rightarrow \chi e^- \gamma$ process.
When the electrons energy is above $E_{res}$, 
the initial state radiation (ISR) diagram takes over,
with a hard ISR photon in the collinear direction that puts
both the internal $e^-$ and $\phi$
propagators on resonance (under the approximation where the electron is massless). 
However, the total cross-section is
suppressed at large incoming 
electron energy, where $\sigma\sim E_{e^-}^{-1}$. As a result, the integrated photon 
spectrum over a power-law spectrum for incoming CR electrons
falls sharply for $E_{\gamma}> M_\phi-M_\chi$.  

For the case of wimp-proton collisions, the up quark's PDF in the proton takes the role of 
reducing the center-of-mass frame energy between the parton and dark matter 
without the necessity of extra radiation. This lifts the
$E_{CR}$ suppression on scattering cross-section at large proton energy. 
\ck{Final state} photons
originate in the hadronic showers around the scattered quark, but more importantly,
from the proton remnants that carry off most of the incoming energy.
However, the PDF preference on relatively low parton momentum fraction suppresses
$\sigma_{\chi p}$ for $E_{p}\sim E_{res}$ and below. The $\sigma_{\chi p}$ shows a steep up-turn 
near $E_{res}$ and continues to increase with $E_p$, which can be seen in the left panel  of Fig.~\ref{fig:e_p_comparison}. Thus there is a larger  contribution of high energy photons compared to the case of dark matter scattering
off electrons. 

The prompt photons in $\chi, p$ collision fall into two major categories: 
\begin{enumerate}
\item Final state radiation (non-remnant FSR). Photons emitted by the hard-scattered u-quark and its shower
belong to this category.
As the proton PDF ensures the s-channel $\phi$ resonance, the final state (non-remnant) energy add up to the
$\phi$ mass, and the FSR photons typically has energy below $\delta M =M_\phi-M_\chi$. However, since $\phi$ and $\chi$
are comparable in mass, the resonance $u-\chi$ system has small  boost and the FSR photons can point
at any direction and dominate the gamma ray signal at large scattering angles.  The shower photons, resulting from hadron decays inside the hadronized jet, have in addition an enhancement due to large multiplicity and the lack of suppression by the QED fine structure constant.  \\

\item Shower from proton remnants.
Similar to the ISR photon in $\chi,e^-$ collisions, at $E_{CR} > E_{res} $ the remnants are emitted along the 
proton's incoming direction. These photons can be emitted at energy much higher than $M_\phi-M_\chi$ but are mainly confined
in this forward region.  Notably,  the number of the photons emitted in proton collisions exceed that in $\chi,e^-$ ISR. At lower incoming
energy $E_{CR} \sim E_{res} $ or less, the collision does not hit resonance. Here, we find that the photons from remnants also
are less energetic than $M_\phi-M_\chi$ but have greater freedom in their direction. 
\end{enumerate}
We calculated the photon spectrum from $\chi,p$ collisions with 
the Monte Carlo generator {\it Sherpa}~\cite{Gleisberg:2008ta}, and
its $Amegic$~\cite{Krauss:2001iv} and $Photons$~\cite{Schonherr:2008av} packages for 
showering and photon radiation. The dark matter Lagrangian is implemented
with the $FeynRules$~\cite{Christensen:2009jx} package.

We use the hadronization model included in {\it Ahadic} \cite{Winter:2003tt} as implemented in {\it Sherpa}. It is designed to allow the study of deep-inelastic scattering (DIS) processes\footnote{We thank Stefan Hoeche for enabling
 Wimps as initial-state particles in the event generation in {\it Sherpa}.}\cite{Bjorken:1969ja}. The resulting partons of our DIS-like hard process, $p \chi \to \chi+X$, and the proton remnants are first showered and then transformed into primordial hadrons during the cluster hadronization process. The subsequent decays of unstable hadrons are also handled by {\it Sherpa}.  In fact, most of the photons produced in the simulation emerge from the decay of neutral mesons, such as $\pi^0$ or through final state radiation in hadron decays.  Admittedly, the modeling for remnants is not experimentally tested for proton collisions with an neutral exotic particle.  Thus, we warn that the theoretical uncertainties on the photon production can be large, especially in the forward region.   

\begin{figure}[h]
\includegraphics[scale=0.6]{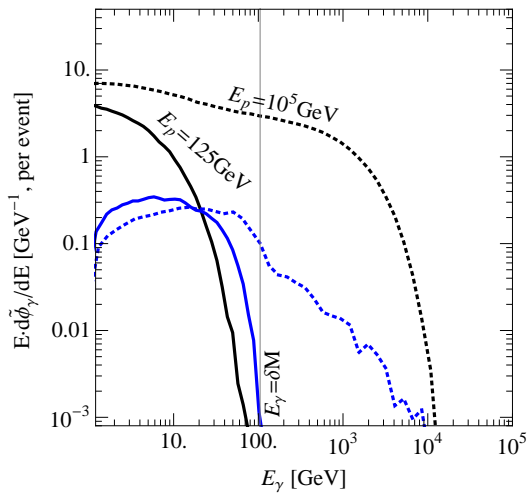}
\caption{
Photon $E\frac{d\phi}{dE}$ spectrum at a near-threshold (125 Gev, solid) and high 
($10^5$ GeV, dashed) proton energy. The spectra for parton level $\chi u\rightarrow \chi u \gamma$  
are shown in blue color, where $\frac{d\phi}{dE}$ is normalized to one photon in each collision.
Black curves are for fully showered $\chi p \rightarrow \chi X$ with remnants  and have 
more photons per event.  For all spectra, the dark matter mass is 300 GeV and the up quark partner is at 405 GeV.  
}
\label{fig:xspec}
\end{figure}

Fig.~\ref{fig:xspec} shows the photon spectrum from one $\chi,p$ collision event, 
comparing the fully showered $\chi u \rightarrow \chi u$ case with the hard photon 
spectrum from a parton level 
calculation of $\chi u\rightarrow \chi u \gamma$.   
For the parton level $\chi u\rightarrow\chi u\gamma$ calculation we imposed these cuts: photon $E,P_T > 1$
 GeV and the invariant mass between $u,\gamma$ greater than 1 GeV. Since the photon bremsstrahlung
has a logarithmic dependence on the charged particle (u-quark) mass, these cuts may cause a 
factor of order ${\cal O}(1)$ to the normalization of parton level radiation, and do not
qualitatively impact the results. 
Here we assume an isotropic distribution
of incoming protons, at 125 GeV that is close to $E_{res}$ (solid curves).  This produces photons mostly below the mass gap $M_\phi-M_\chi$.  This should be contrasted with the high energy case of $E_p=10^5$ GeV (dashed curves.  
Note that the gap between the fully showered and parton level
spectra widens for higher energy incoming protons, due to extra hard photons from remnants. For $E_p$ above resonance energy, 
while the FSR photons remain below $M_\phi-M_\chi$, ISR and proton remnants photons
continue  to higher energies. 

\begin{figure}
\includegraphics[scale=0.6]{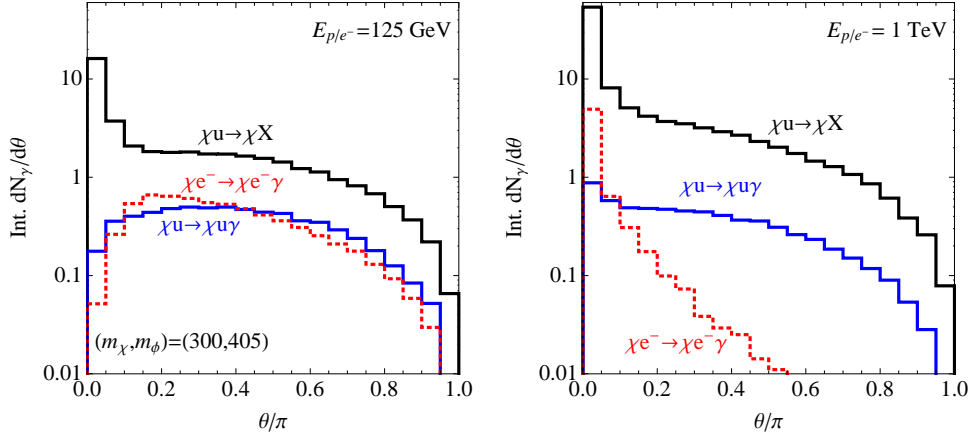}
\caption{
Angular distribution of photons with $E_\gamma >1$ GeV at cosmic ray energy near (left panel) and above (right panel)
the resonance energy. 
$\theta$ denotes the photon's scattering angle off the incoming proton direction.
Showered spectra are normalized to the 
number of final state photons, $\tilde{N}_\gamma$ (above 1 GeV). 
Parton level curves
are normalized to one photon per event and their gap to the fully 
showered curve at large scattering angle is due to $\tilde{N}_\gamma$.
Note that above $E_{res}$ the ISR dominates the $\chi e^-\rightarrow \chi e^-\gamma$
spectrum and emission at large angle diminishes. Furthermore, the behavior at small $\theta$ show how proton remnants provide extra forward photons.
For incoming particles near resonance energy (left), photons from hard event are less peaked in the
forward direction. The $\chi e^-$ collision results are also shown for comparison.  For all spectra, the dark matter mass is 300 GeV and the \ck{up quark (electron)} partner is at 405 GeV.  The parton level spectra have kinematic cuts
that require both photon $P_T$ and $u/e^-,\gamma$ invariant mass above 1 GeV.
 }
\label{fig:ang_dists}
\end{figure}

In Fig.~\ref{fig:ang_dists}, we  show the distribution of photons ($E_\gamma > 1$ GeV) over the scattering angle. 
Here the scattering angle $\theta$ is between the photon and the incoming proton's direction.
Note that the remnants give a more pronounced peak in the forward direction for $\chi,p$ collisions, 
as exemplified by the 1 TeV incoming proton. Although
$\chi e^-$ collisions for  high incoming electron energy also favors a forward ISR photon, 
the distribution of $\chi,e^-$
signal is dominated by $\sigma_{\chi e^-}$ near $E_{res}$, as seen in left panel of Fig.~\ref{fig:e_p_comparison}.

\begin{figure}[h]
\includegraphics[scale=0.6]{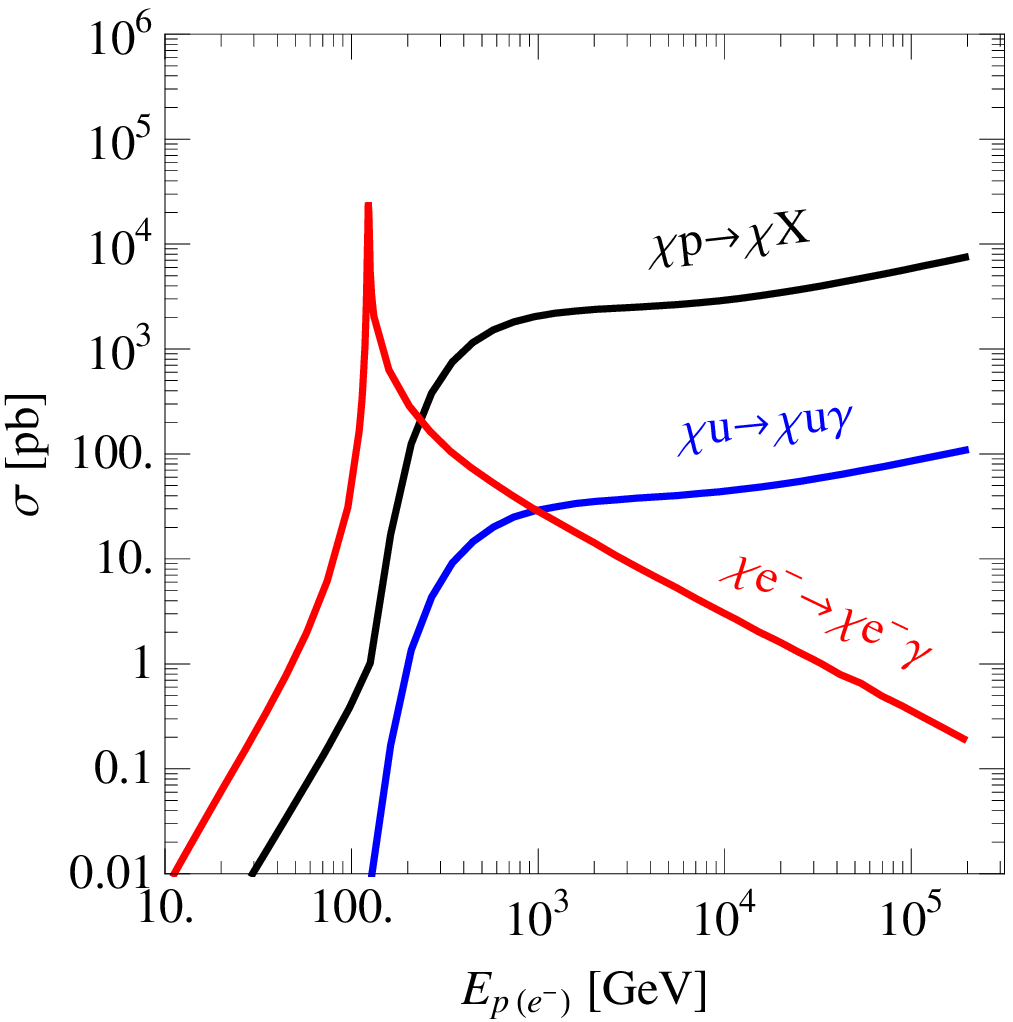}
\includegraphics[scale=0.6]{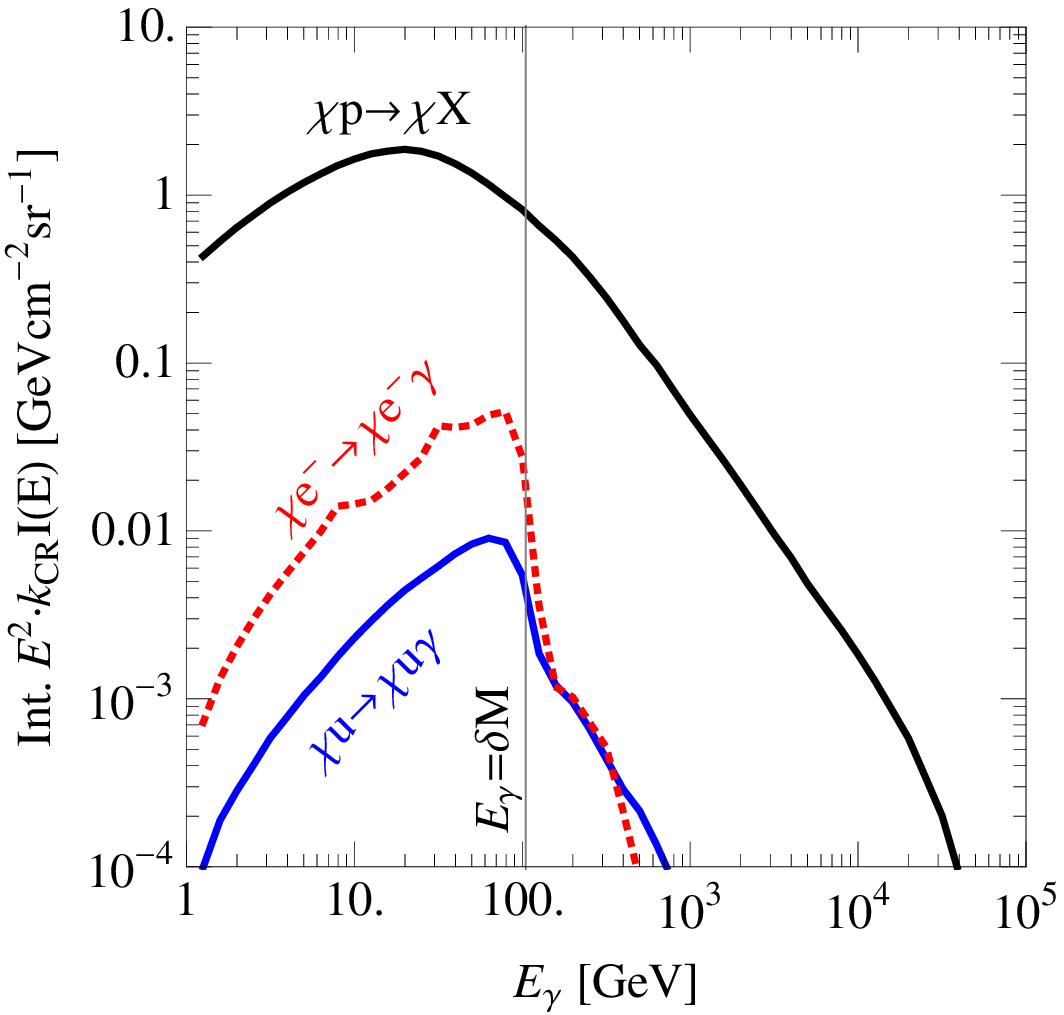}
\caption{
(left:) Dark matter - cosmic $p/e^-$ 
scattering cross-sections. The parton-level $\chi u\rightarrow \chi u \gamma$ cross-section
depends on the kinematic cut on the final state photon, besides a generic $\alpha_{QED}$
suppression in comparison to that of the leading order $\chi u\rightarrow \chi u$.
(right:) Unit-volume gamma ray spectra from full proton shower, 
parton level $\chi u \rightarrow \chi u \gamma$ and 
$\chi e^- \rightarrow \chi e^- \gamma$. 
In the right panel, the integrated flux 
$I(E_\gamma)$ is given in Eq.~\ref{eq:energy_integral} with a reference
energy $E_0=10$ GeV. The CR flux is taken
as Eq.~\ref{eq:CR_flux} with normalizations $k_p=1$ and $k_{e^-}=0.01$.
In both panels the $\sigma_{\chi e^-}$ is scaled up by a factor of $\sim$3 
to compensate for the kinematic cuts.  For all spectra, the dark matter mass is 300 GeV and the \ck{up quark (electron)} partner is at 405 GeV.  
}
\label{fig:e_p_comparison}
\end{figure}

The cross-sections for $\chi, p$ scattering are shown in Fig.~\ref{fig:e_p_comparison}.  Scattering with
an electron is also plotted as a comparison for the high energy behavior of the cross section.
The 
$\chi (u/e^-) \rightarrow \chi (u/e^-) \gamma$ processes 
use the same set of cuts as in Fig.~\ref{fig:xspec}.
In the right panel of Fig.~\ref{fig:e_p_comparison}, 
the photon spectra from dark matter collisions with galactic cosmic rays is generated by convolving with a $E^{-2.7}$
proton spectrum ($E^{-3}$ for $e^-$) that is typical for the cosmic rays in the central region
of our Galaxy. The normalization on the electron flux is $10^{-2}$ of that of protons; 
see Appendix ~\ref{app:source_spec}
for the definition of $I(E)$. Due to the lack of peaking near $E_{res}$, $\chi,p^-$ scattering 
yields a harder gamma ray signal above $M_\phi-M_\chi$.  Compared to the parton level $\chi u\rightarrow \chi u\gamma$ prediction,  we see that the showers
give a roughly two order of magnitude enhancement in photons for proton-dark matter scattering.  As previously mentioned, this difference  is due to
the $\alpha_{QED}$ suppression of the parton level cross section and a 
higher final state photon multiplicity $\tilde{N}_\gamma$
from full hadronization/showering.

\section{The AGN case: Centaurus A}
\label{sect:AGN}

An interesting place to look for $\chi p$ collision is at the center of nearby AGNs,
where dark matter halo is assumed to exist and luminous jets provide high energy protons. 
We take Centaurus A for our calculations.
Although the proton composition of the AGN jet has large modeling uncertainty,
protons can make up a majority of jet particles and the energy output in protons, $L_p$, from Cen.~A can be 
more than $10$ times higher than in leptons~\cite{Falcone:2010fk}.
In this section, we compute the gamma ray signal arising from proton-dark matter collisions, 
with a focus on the contribution from proton remnants. 

We make an assumption that protons are isotropic in the AGN's `blob' frame, similar to the jet electrons.
The spectrum of protons are not well known.  We assume that the protons also
undergo Fermi acceleration and their isotropic spectrum in
the `blob' frame is a power-law $E^{-s}$, where the index $s = 2$:
\bea  
\frac{d\dot{N}_p}{d\tilde{E}d\tilde{\Omega}}&=& \frac{K_p}{4\pi}\left(\frac{\tilde{E}}{E_0}\right)^{-s}  
\label{eq:AGN_powerlaw}
\eea
The tilde $\tilde{\ }$ denotes variables in the boosted `blob' frame. $K_p$ is the normalization
that is determined by AGN's proton output.
The `blob' frame moves at a Lorentz boost factor $\Gamma_B$ relative to the
central black hole.  Following the analyses in \cite{Gorchtein:2010xa, Huang:2011dg}, we take $\Gamma_B = 3$.
Boosting back to the black hole frame where the 
wimps are non-relativistic, the proton spectrum is
\be
\frac{d\dot{N}_p}{d{E}d\Omega} 
=\frac{d\dot{N}_p}{d\tilde{E}d\tilde{\Omega}}\cdot \frac{1}{\Gamma_B(1-\beta_B \cos\theta)} 
\label{eq:jacobian}
\ee
where the $\beta_B = 0.94$ is the `blob' frame's velocity. The energy and zenith angle before/after the boost are related by
\be 
\cos\tilde{\theta}=\frac{\cos\theta-\beta_B}{1-\beta_B \cos\theta} \hspace{1cm} \tilde{E}=E\cdot\Gamma_B(1-\beta_B \cos\theta).
\label{eq:variable_change}
\ee
with $\theta=0$ along the jet axis. 
Combining Eq.~\ref{eq:jacobian} and~\ref{eq:variable_change}, the black-hole frame proton spectrum is
\be 
\frac{1}{2\pi}\frac{d\dot{N}_p}{dE~d\cos\theta}=\frac{K_p}{4\pi} \left(\frac{E}{E_0}\right)^{-s}
\cdot \left[\Gamma_B~(1-\beta_B\cos\theta)\right]^{-(s+1)},
\label{eq:BHspectrum}
\ee
which is still a power-law and has the same index as that in the `blob' frame, 
while its intensity now varies with direction. For relativistic particles,
the reference energy $E_0$ is irrelevant and can be absorbed into the normalization.  
Keeping a nonzero mass leads to
a ${\cal O}(\gamma_{p}^{-2})$ correction to the formulae above, \ck{where $\gamma_{p}= E/m_p$.}
Integrating Eq.~\ref{eq:BHspectrum} with proton energy gives the
total proton energy output as
\bea 
L_p&=&
K_p\cdot \frac{E_0^s}{2s(2-s)\beta_B\Gamma_B^{s+1}}
\left(
E_{max}^{2-s}-E_{min}^{2-s}
\right)
\left[
(1-\beta_B)^{-s}-(1+\beta_B)^{-s}
\right],\\
\text{or}~K_p&=&\frac{L_p}{E_0^{2}~\Gamma_B~\ln(E_{max}/E_{min})},~\text{for }s=2.
\eea

As Eq.~\ref{eq:BHspectrum} shows, while the flux along the jet direction is greatly enhanced, 
the Jacobian suppression at large
angles $\propto \Gamma_B^{-3}$.  
The Cen.~A jet is $68\degree$ ($\cos\theta=0.37$) from the Earth;
in this  direction, the Jacobian from the Lorentz boost suppresses 
the proton flux by a factor of 0.14 compared to the unboosted flux.
Most protons are along the jet axis, thus their gamma ray contribution towards the Earth is
through large angle scattering. In terms of phase-space, for these `along-axis' protons,
a $4\pi$ integration of the Jacobian $\int \left|J\right| \text{d}\Omega$ is of order ${\cal O}(10^2)$ times
favored by the Lorentz boost, in comparison to the Jacobian integrated around an angular window $\Delta\cos\theta<0.1$
centered on the protons pointed towards the Earth.
Given the significant
enhancement from proton remnants, 
the  ${\cal O}(10-10^2)$ higher photon flux 
can negate/overcome this suppression at large incoming proton energy. While the low $E_\gamma$
spectrum is still dominanted by radiation from the protons along the jet axis, 
protons that point near to the Earth also make considerable contribution to the gamma ray signal,
especially when $E_\gamma>M_\phi-M_\chi$.
 
Admittedly the Lorentz boosted `blob frame' is a simplistic picture for the protons inside
the AGN jet. If the AGN jet is more collimated than our assumption, less protons would point towards the Earth
and the photons in the forward region would play a less important role.  To illustrate this uncertainty in the AGN proton distribution, our signal prediction will also be shown with just the `along-axis' protons with $\cos \theta_p >0.8$.
 

Including radiation from protons in all directions, the photon flux towards Earth is
\be
\left.\frac{d\phi_\gamma}{dE_\gamma d\Omega}\right|_{\vec{\theta}_\oplus}=\frac{1}{R^2}
\frac{\delta_{DM}}{M_\chi}
\int d\Omega_p
\int dE_p ~\sigma(E_p)\frac{dN_p}{dE_p d\Omega_p}
\left.\frac{d\tilde{N_\gamma}}{dE_\gamma d\Omega_{\gamma, sc}}
\right|_{\theta_{\gamma,sc}=<\vec{\theta}_p,\vec{\theta}_\oplus >} .
\label{eq:AGN_sig}
\ee
Throughout this paper we denote angular-integrated cosmic ray flux as $\phi$ and its angular differential form as
${d\phi}/{d\Omega}$.
The direction of the incoming proton $\vec{\theta}_p=\{\theta_p,\phi_p \}$ is not limited to the vicinity of the jet axis.
$\theta_{\gamma, sc}$ denotes the `real' photon scattering angle in a frame where  
the proton momentum is along the $z$-axis. $\theta_{\gamma, sc}$ is determined by the proton direction 
$\vec{\theta}_p$ and the Earth's direction $\vec{\theta}_\oplus$.
\ck{$R=3.7$~Mpc is the Earth's distance to Cen. A}.
The integrand in Eq.~\ref{eq:AGN_sig} determines the contribution from protons at angle $\theta_p$ off the AGN jet axis.
$\delta_{DM}=\left< \rho_{\chi}(r)\cdot r \right>$ is the dark matter halo density integrated over the distance range
where collisions occur; $\delta_{DM}$ at Cen. A can be as high as $10^{11} M_\odot/$pc$^2$~\cite{Gorchtein:2010xa}.
For \ck{the jet output in protons}, we use $L_p=$\sci{1}{45} erg~s$^{-1}$ \ck{and an energy range
$[E_{min},E_{max}]=[10,10^7]$ GeV}~\cite{Falcone:2010fk}. $\frac{d\tilde{N}}{d E_{\gamma}d\Omega_{\gamma,sc}}$ 
denotes the final state photon distribution from an average collision event. $\frac{d\tilde{N}}{d E_{\gamma}d\Omega_{\gamma,sc}}$  is normalized 
to the total number of photons above 1 GeV per collision and is Monte Carlo generated.  

\begin{figure}[h]
\includegraphics[scale=.7]{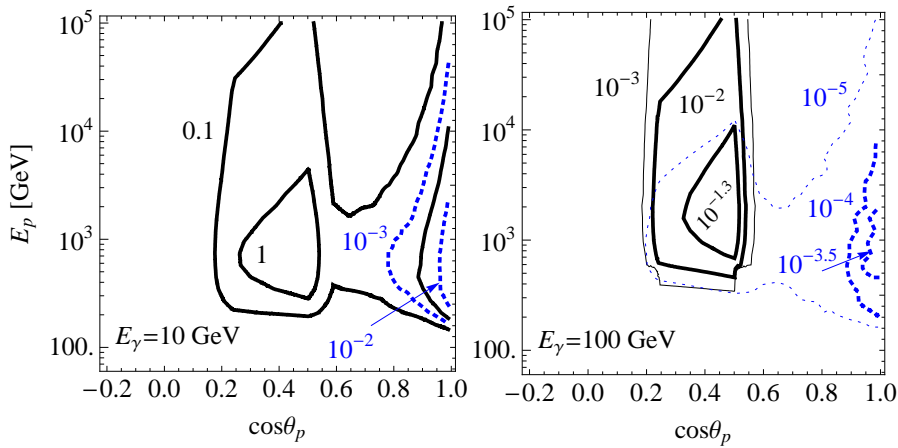}
\caption{
Emission intensity contours in terms of proton energy and proton's angle 
off the jet axis. Black and blue
contours denote the showered (solid) and parton level (dotted) calculations, respectively.
The azimuthal angle $\phi_p=0$ in both panels, i.e. protons are in the Earth - jet axis plane.
The Earth's direction \ck{is} $\cos \theta_p$=0.37. Emission intensity $f$ is the integrand in Eq.~\ref{eq:AGN_sig},
in units of pb$\cdot$GeV$^{-2}\cdot$sr$^{-2}\cdot\left.K_p\right|_{E_0=10~\text{GeV}}$.
For all spectra, the dark matter mass is 300 GeV and the  up quark partner is at 405 GeV.
}
\label{fig:AGN_sig_contour}
\end{figure}

\begin{figure}[h]
\includegraphics[scale=0.7]{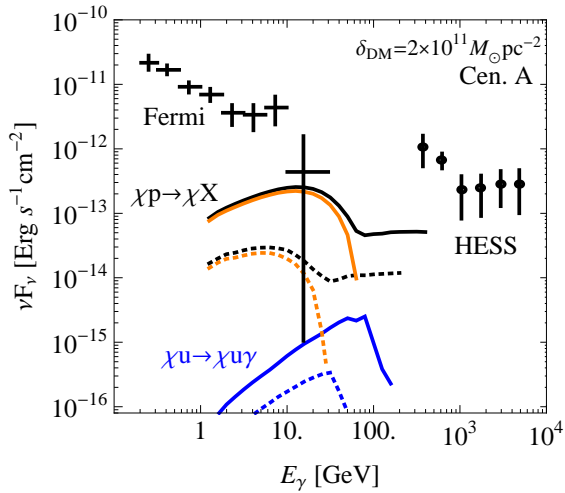}
\caption{
$\chi,p$ collision induced gamma ray signal from
Cen. A, for sample point A (solid) and B (dotted).
The full collision and parton level $E^2\frac{d\phi}{dE}$ spectra are plotted in black and blue colors, respectively.
The orange curves show the component of the fully showered spectra that originate from protons along the AGN jet axis, $\cos \theta_p >0.8$.
Signal levels assume optimistic AGN parameters: $L_p=10^{45}$erg~s$^{-1}$ and 
$\delta_{DM}=2\times 10^{11} M_\odot$pc$^{-2}$~\cite{Gorchtein:2010xa}. 
Fermi~\cite{Falcone:2010fk} and Hess~\cite{Raue:2009vp} measurements are shown for comparison.
 }
\label{fig:cenA}
\end{figure}

As shown in Fig.~\ref{fig:AGN_sig_contour}, the protons along the jet axis ($\cos\theta_p \sim 1$) 
suffices for parton level radiation (shown in blue dotted contours).  The showered spectra (shown in black solid contours) shows that 
the forward photons, with $\cos \theta_p \sim 0.37 $, mostly from proton remnants, account for a significant portion or even the 
majority of the signal, especially at large $E_\gamma$.  Even at low $E_\gamma$, contributions from 
protons along the jet axis are still significant.
The resulting gamma ray signal at the Earth is plotted in Fig.~\ref{fig:cenA} in black.  The signal is noticeably enhanced over the parton-level calculation shown in blue.  Furthermore, the shape is substantially altered.  At low energies, $E_{\gamma}<M_\phi-M_\chi$, the spectra is softer than the parton-level result and has less of a peaking structure.     
At high energies, $E_{\gamma} > M_\phi-M_\chi$, the fully showered
spectrum is a power law at high energy which  receives a
significant contribution from protons along the jet axis. 
In comparison, the parton level photons drop abruptly after 
reaching $M_\phi-M_\chi$. Thus, taking into hadronization and showering 
has both significantly enhanced the signal and altered its spectral shape. 
Note that due to numeric stability issues in Monte Carlo, we do not plot $E_\gamma$ above
1 TeV, yet the $E^{-2}$ power-law spectrum is expected to extend to higher energy, as a fragmentation 
from the the total incoming proton energy.

In  Fig.~\ref{fig:cenA} , we have also plotted the signal component from the protons that are just along the AGN jet axis (shown in orange),  integrating over  $\cos\theta_p>0.8$ in Eq.~\ref{eq:AGN_sig}.  This demonstrates the situation for a more highly collimated AGN jet, where only  large-angle 
scattering gamma rays contribute.  Thus, the high energy tail above $M_\phi-M_\chi$ is sensitive to the theoretical uncertainties of the AGN jet angular distribution as well as  the photon contribution from proton remnants.   As these curves show, for large angle scattering, the  major difference is that such photons drop much
more abruptly when $E_\gamma$ approaches to $M_\phi-M_\chi$. The parton-level curves, while their normalizations
are suppressed by $\alpha_{QED}/\tilde{N}_\gamma$, also demonstrate this high energy behavior of photons from hard scattering.

The overall signal level scales linearly with $\delta_{DM}$ and the AGN's energy output in protons.
In illustrating the gamma ray signal we assumed an optimistic scenario 
with regards to the values of the dark matter density, the AGN's proton energy output and the interaction coupling. 
The resulting gamma ray signal level for sample point A is comparable to the uncertainties in the Fermi data
and future observation may constrain the coupling to lower values.  
More optimistically, with further enhancements to the dark matter signal, 
the high energy tail could explain the HESS data points without 
modifying the lower energy Fermi points and thus resolve the discrepancy 
in power law observed by the HESS and Fermi-LAT analyses~\cite{Falcone:2010fk}.

\section{Diffuse protons}
\label{sect:diffuse}

For an isotropic distribution of protons, e.g. the diffuse protons inside the Milky Way, 
there is no prefered direction and the contribution from proton remnants are present
in the $4\pi$-averaged prompt spectrum. However, due to relatively low CR flux inside the Milky Way
plus a high energy threshold for resonance scattering,
the gamma ray signal is much below galactic background levels. 
In this section we only describe the calculations with two template profiles of
galactic protons.

The photon signal is given by,
\be 
\frac{d\phi_\gamma}{dE_\gamma}=\int d{r}\frac{\rho_{\chi}({r})}{M_\chi}
\int_{E_{\gamma}}^{+\infty} dE_{p} \sigma_{\chi p}\frac{d\phi_p}{d E_p}\frac{d\tilde{N}}{dE_\gamma}
\label{eq:int_spec}
\ee
where $\rho({\bf r})$ is the dark matter halo density, $\frac{d \phi_p}{d E_p}$
is the cosmic ray flux. At the center of the Milky Way galaxy, these
fluxes can be parametrized~\cite{Cirelli:2010xx} as
\bea 
\frac{d \phi_p}{d E_p d\Omega}&=& k_p\left(\frac{E_p}{\text{GeV}}\right)^{-2.7}
\text{GeV}^{-1}\text{cm}^{-2}\text{s}^{-1}\text{sr}^{-1}\hspace{1cm}\text{for protons} \nn \\
 \frac{d \phi_{e^-}}{d E_{e^-}d\Omega}&=& k_{e^-}\left(\frac{E_{e^-}}{\text{GeV}}\right)^{-3}
\text{GeV}^{-1}\text{cm}^{-2}\text{s}^{-1}\text{sr}^{-1}\hspace{.6cm}\text{for electrons}
\label{eq:CR_flux}
\eea
Note: the flux normalizations $k_{p/e^-}$ in lower-case are not to be confused with that of the AGN
jet.
Since the diffuse spectrum is isotropic, the forward photons are readily present 
and their spectrum is illustrated in Fig.~\ref{fig:xspec}.

For protons, its power law index only varies slightly during 
propagation and the spatial and energy
parts in  Eq.~\ref{eq:int_spec} can be separated,
\be 
\frac{d\phi_{\gamma}}{d E_{\gamma}d\Omega}({\bf \theta})=
{ J}({\bf \theta})\cdot  I(E_\gamma)
\ee
where ${J(\theta)}$ integrates over the dark matter distribution along
the direction ${\bf \theta}$, while $I(E_\gamma)$ is the prompt gamma spectrum
convoluted with the proton energy spectrum.
For details see Appendix~\ref{app:source_spec}.

\begin{table}[h]
\begin{tabular}{c|c|c|c}
\hline
CR model	&\ \  $\alpha_p$\ \  &	\ \ $M_\chi\cdot\bar{J}(\theta)$ Central\ \ 	& \ \ $M_\chi\cdot\bar{J}(\theta)$ Inner\ \ \\
\hline
Plain diffusion~\cite{Ptuskin:2005ax}	& -2.68 & 3.8	&	0.65	\\
Diffusion reacc.~\cite{bib:reacc} & -2.75 & 9.9	&	1.5	\\
\hline
\end{tabular}
\caption{
\ck{Angular averaged $M_\chi\cdot \bar{J}$ in central ($|\theta|< 1\degree$ ) and inner 
($|l|<80\degree,|b|<8\degree$) galactic regions. $\alpha_p$ is the power index of the proton flux. 
$J$ is evaluated with reference energy $E_0$ at 10 GeV. See Eq.~\ref{eq:def_J} for definitions.
The dark matter profile is given in Eq.~\ref{eq:enaisto}.
$M_\chi\cdot J(\theta)$ values are in the unit of $10^{28} $s$^{-1}$m$^{-4}$sr$^{-1}.$}}
\label{tab:int_J}
\end{table}

We choose two template Galprop CR \ck{profiles}, the plain diffusion model 999726~\cite{Ptuskin:2005ax} 
and diffusion-reacceleration model 599278~\cite{bib:reacc} to calculate the integrated strength 
of gamma ray source, $\int J(\theta) d\theta$ in Fermi's angular 
windows~\cite{FermiLAT:2012aa}, as shown in Tab.~\ref{tab:int_J}. For the dark matter
halo we pick the Einasto profile~\cite{Navarro:2008kc} as an example for cuspy dark matter distribution,
\be 
\rho_\chi=\rho_{\odot}\text{e}^{-\frac{2}{\alpha}[(r^{\alpha}-r_{\odot}^\alpha)/r_s^\alpha] },
\hspace{1cm} 
\label{eq:enaisto}
\ee
where $\alpha=1.7$, $r_\odot=8.3, r_s=25$~kpc and the local halo density $\rho_\odot=0.3$ GeV/cm$^3$.

Inside the Milky Way, however, the proton-dark matter scattering is at an disadvantage to due the relatively
low level of cosmic ray flux. Small mass splitting between $\phi, \chi$ may face increasingly stringent
constraint from direct detection experiments.
As Eq.~\ref{eq:res} and Fig.~\ref{fig:masses} illustrate, only cosmic protons
of \ck{${\cal O}(10-100)$} GeV or above significantly contribute to the gamma ray signal for our toy Lagrangian.
As the result the $p,\chi$ collision
signal is dwarfed in comparison to the astrophysical background.

\section{Summary}
\label{sect:summary}

In this paper we investigated the gamma ray signal from the collision between dark matter 
and high energy cosmic ray protons, including the effects of hadronization and showering.  
This extends previous parton-level only calculations, with a substantial modification of the rate and energy spectrum of the photons.
The rate is significantly enhanced, since the photons produced in hadronic decays have a high multiplicity and are not suppressed by the fine structure constant.
In particular, we emphasize the contribution
from the energetic proton remnants, which boost the high energy tail of the gamma ray spectra.  To illustrate the shower enhancement 
to a parton level photon radiation, we implemented a simple Majorana
fermion dark matter that couples right-handedly to the up-quark, to avoid
large annihilation rates. We used the latest XENON100 limits to select
viable test masses for the dark matter and scalar $u$ partner for the
cosmic signal from Cen A as an AGN candidate, and the case of diffuse
protons in our Milky Way galaxy.

We use the Monte Carlo generator Sherpa to simulate one-sided proton 
remnant in a $\chi, p$ collision event and subsequent showering.
Due to limited choice of generators that allow remnant showering with 
exotic particle beams, it is of interest to further test 
the photon radiation from proton remnants with alternative means of 
calculation.  This will help to determine whether there are large theory uncertainties on this gamma ray signal.
An additional source of uncertainty is the amount of AGN protons which are pointed towards the Earth.  This also affects the high energy photons and thus, improvements in AGN modeling will also help pin down this part of the spectrum.

To summarize, we find that hadronization and showering substantially enhance the signal and in particular, the proton remnants significantly  enhance the signal rate for the energy range $E_\gamma > M_\phi-M_\chi$, making the protons pointed directly
 towards the Earth the major contributor at large photon energy.
The gamma ray signals from proton-dark matter collision is found to be at a 
level which could be potentially constrained by forthcoming Fermi measurements. In contrast,
collisions with diffuse protons inside the Milky Way suffers from the
low proton flux at energies high enough to reach
resonance scattering. However, if the dark matter candidate and the mediator are light the signal from diffuse protons may become more prominent. \\

{\bf \noindent Acknowledgements}
\label{acknowledgements}
We thank Stefan Hoeche for his help with exotic beams in Sherpa, \ck{and Chris Savage for providing codes that implement
the spin dependent form factor.} We also thank Frank Kraus, Tim Tait and Jinrui Huang for helpful discussions. This paper is supported by DOE under grant \# DE-FG02-96ER40969.

\appendix
\section{Diffuse cosmic ray scattering}
\label{app:source_spec}

The prompt photon flux at the Earth is an integral over collision sources inside the observation 
angular cone $\Delta\Omega$,
\bea
\frac{d \phi_\gamma}{dE_\gamma}&=& \int \frac{dV}{4\pi r^2} \frac{dN_\gamma}{dE_\gamma dV} \nn \\
&=&\int \frac{d{\bf r}}{4\pi |{\bf r}|^2}\cdot 
\hspace{0.2cm} 4\pi\int_{E_\gamma}^{+\infty} \frac{\rho_{\chi}({\bf r})}{M_\chi}\frac{d\phi_p}{dE_p}({\bf r})\frac{d\sigma(E_p,E_\gamma)}{dE_\gamma} 
\label{eq:full_spec_int}
\eea
where ${\bf r}=(r,\boldmath\theta)$ with {$\theta$} 
in galactic coordinates, $\rho_{\chi}$ is the dark matter density and
$\frac{d\phi_p}{dE_p}$ is the $4\pi$ averaged galactic proton flux.

Only the high-energy protons are relevant in our study and the spectrum
can be parametrized as a power-law,
\be 
\frac{d\phi_p}{dE_p}=\phi^0_p(E_0)\left(\frac{E_p}{E_0}\right)^{-s}\hspace{0.5cm}\text{for $E>E_0$}
\ee
where the power index $s$ grows slightly as protons lose energy during their propagation
to the outer region of the galaxy.
In the Galprop models we adopted, the variation
$\delta s\sim 10^{-2}$ in the power-law index is insignificant and $s$ can be approximated as a constant. 

Thus the energy integral can be separated from the spatial one and Eq.~\ref{eq:full_spec_int}
can be written into
\bea 
\frac{d\phi_\gamma}{dE_\gamma d\Omega}&=& J({\bf \theta}) \cdot I(E_\gamma), \label{eq:JI_separation}\\
\text{where}\hspace{1cm} J(\mathbf{\theta})&=&
     \int_{0}^{+\infty} dr \frac{\rho_\chi(r,\theta)}{M_\chi} \phi^0_p(E_0,r,\theta), 
     \label{eq:def_J}\\
I(E_\gamma)&=&\int_{E_\gamma}^{+\infty} dE_p\left(\frac{E_p}
     {E_0}\right)^{-s}\frac{d\sigma(E_p,E_\gamma)}{dE_\gamma}. \label{eq:energy_integral}
\eea
The energy integral $I(E_\gamma)$ gives the shape of the prompt gamma ray spectrum and 
is independent from astrophysics. In the case with hadronic shower, the differential cross-section
in Eq.~\ref{eq:energy_integral} is replaced with
\be 
\frac{d\sigma(E_p,E_\gamma)}{dE_\gamma}\equiv \sigma_{tot}(E_p)\frac{d\tilde{N}_\gamma}{dE_\gamma}.
\ee


\begin{thebibliography}{99}
\label{bibs}

\bibitem{Falcone:2010fk} 
  A.~A.~Abdo {\it et al.}  [Fermi Collaboration],
  Astrophys.\ J.\  {\bf 719}, 1433 (2010)
  [arXiv:1006.5463 [astro-ph.HE]].


\bibitem{Raue:2009vp} 
  M.~Raue {\it et al.}  [H.E.S.S. Collaboration],
  arXiv:0904.2654 [astro-ph.CO].

\bibitem{Bloom:1997vm} 
  E.~D.~Bloom and J.~D.~Wells,
  Phys.\ Rev.\ D {\bf 57}, 1299 (1998)
  [astro-ph/9706085].

\bibitem{Gorchtein:2010xa}
  M.~Gorchtein, S.~Profumo and L.~Ubaldi,
  Phys.\ Rev.\ D {\bf 82} (2010) 083514
   [Erratum-ibid.\ D {\bf 84} (2011) 069903]
  [arXiv:1008.2230 [astro-ph.HE]].
  
\bibitem{Huang:2011dg} 
  J.~Huang, A.~Rajaraman and T.~M.~P.~Tait,
  arXiv:1109.2587 [hep-ph].


  
\bibitem{Profumo:2011jt} 
  S.~Profumo and L.~Ubaldi,
  JCAP {\bf 1108}, 020 (2011)
  [arXiv:1106.4568 [hep-ph]].


\bibitem{Hisano:2011um} 
  J.~Hisano, K.~Ishiwata and N.~Nagata,
  Phys.\ Lett.\ B {\bf 706}, 208 (2011)
  [arXiv:1110.3719 [hep-ph]].

\bibitem{Aprile:2012nq} 
  E.~Aprile {\it et al.}  [XENON100 Collaboration],
  arXiv:1207.5988 [astro-ph.CO].

\bibitem{Drees:1993bu} 
  M.~Drees and M.~Nojiri,
  Phys.\ Rev.\ D {\bf 48}, 3483 (1993)
  [hep-ph/9307208].

\bibitem{Hisano:2010ct} 
  J.~Hisano, K.~Ishiwata and N.~Nagata,
  Phys.\ Rev.\ D {\bf 82}, 115007 (2010)
  [arXiv:1007.2601 [hep-ph]].


\bibitem{Adriani:2010rc} 
  O.~Adriani {\it et al.}  [PAMELA Collaboration],
  Phys.\ Rev.\ Lett.\  {\bf 105}, 121101 (2010)
  [arXiv:1007.0821 [astro-ph.HE]].

\bibitem{lightdm}
C.~Boehm and P.~Fayet,
  Nucl.\ Phys.\ B {\bf 683}, 219 (2004),
 [hep-ph/0305261];
G.~Belanger, S.~Biswas, C.~Boehm and B.~Mukopadyaya,
  arXiv:1206.5404 [hep-ph].

\bibitem{Gleisberg:2008ta} 
  T.~Gleisberg, S.~.Hoeche, F.~Krauss, M.~Schonherr, S.~Schumann, F.~Siegert and J.~Winter,
  JHEP {\bf 0902}, 007 (2009)
  [arXiv:0811.4622 [hep-ph]].
  
\bibitem{Krauss:2001iv}
  F.~Krauss, R.~Kuhn and G.~Soff,
  JHEP {\bf 0202} (2002) 044
  [arXiv:hep-ph/0109036].

\bibitem{Schonherr:2008av} 
  M.~Schonherr and F.~Krauss,
  JHEP {\bf 0812}, 018 (2008)
  [arXiv:0810.5071 [hep-ph]].


\bibitem{Christensen:2009jx}
  N.~D.~Christensen, P.~de Aquino, C.~Degrande, C.~Duhr, B.~Fuks, M.~Herquet, F.~Maltoni and S.~Schumann,
  Eur.\ Phys.\ J.\ C {\bf 71} (2011) 1541
  [arXiv:0906.2474 [hep-ph]].
 
 \bibitem{Winter:2003tt} 
  J.~-C.~Winter, F.~Krauss and G.~Soff,
  Eur.\ Phys.\ J.\ C {\bf 36}, 381 (2004).
 [hep-ph/0311085].
 
 \bibitem{Bjorken:1969ja} 
  J.~D.~Bjorken and E.~A.~Paschos,
  Phys.\ Rev.\  {\bf 185}, 1975 (1969).
 
\bibitem{Cirelli:2010xx}
  M.~Cirelli, G.~Corcella, A.~Hektor, G.~Hutsi, M.~Kadastik, P.~Panci, M.~Raidal and F.~Sala {\it et al.},
  JCAP {\bf 1103} (2011) 051
  [arXiv:1012.4515 [hep-ph]].

\bibitem{Ptuskin:2005ax} 
  V.~S.~Ptuskin, I.~V.~Moskalenko, F.~C.~Jones, A.~W.~Strong and V.~N.~Zirakashvili,
  Astrophys.\ J.\  {\bf 642}, 902 (2006)
  [astro-ph/0510335].
  
\bibitem{bib:reacc} 
Wiedenbeck, M.~E., 
Yanasak, N.~E., Cummings, A.~C., et al.\ 2001, Space Science Reviews, 99, 15 



\bibitem{FermiLAT:2012aa} 
  [Fermi-LAT Collaboration],
  Astrophys.\ J.\  {\bf 750}, 3 (2012)
  [arXiv:1202.4039 [astro-ph.HE]].
  
\bibitem{Navarro:2008kc} 
  J.~F.~Navarro, A.~Ludlow, V.~Springel, J.~Wang, M.~Vogelsberger, S.~D.~M.~White, A.~Jenkins and C.~S.~Frenk {\it et al.},
  arXiv:0810.1522 [astro-ph].

\end{thebibliography}
\end{document}